# Not Only Because of Theory: Dyson, Eddington and the Competing Myths of the 1919 Eclipse Expedition

## Introduction

One of the most celebrated physics experiments of the 20th century, a century of many great breakthroughs in physics, took place on May 29th, 1919 in two remote equatorial locations. One was the town of Sobral in northern Brazil, the other the island of Principe off the west coast of Africa. The experiment in question concerned the problem of whether light rays are deflected by gravitational forces, and took the form of astrometric observations of the positions of stars near the Sun during a total solar eclipse. The expedition to observe the eclipse proved to be one of those infrequent, but recurring, moments when astronomical observations have overthrown the foundations of physics. In this case it helped replace Newton's Law of Gravity with Einstein's theory of General Relativity as the generally accepted fundamental theory of gravity. It also became, almost immediately, one of those uncommon occasions when a scientific endeavor captures and holds the attention of the public throughout the world.

In recent decades, however, questions have been raised about possible bias and poor judgment in the analysis of the data taken on that famous day. It has been alleged that the best known astronomer involved in the expedition, Arthur Stanley Eddington, was so sure beforehand that the results would vindicate Einstein's theory that, for unjustifiable reasons, he threw out some of the data which did not agree with his preconceptions. This story, that there was something scientifically fishy about one of the most famous examples of an *experimentum crucis* in the history of science, has now become well known, both amongst scientists and laypeople interested in science.

Yet this story has hardly ever been itself subjected to a close examination. It is the contention of this essay that there are no grounds whatsoever for believing that personal bias played any sinister role in the analysis of the eclipse data. Furthermore there are excellent grounds for believing that the central contention made by the expedition's scientists (including Eddington), namely that the results were roughly consistent with the prediction of Einstein's theory of General Relativity and firmly ruled out the only other theoretically predicted values (including the so-called "Newtonian" result), was indeed justified by the observations taken.

## Overview

The basic outline of the story that I wish to rebut, as with most compelling narratives, is simple. Arthur Stanley Eddington fervently wished for a confirmation of general relativity for two reasons. He was a firm believer in and advocate of the theory, and was utterly convinced that the prediction of light-bending it made was true. He was a pacifist and war resister who earnestly sought postwar reconciliation between scientists in Britain and Germany, and saw the confirmation by a British expedition of the theory of

Germany's leading scientist as a heaven-sent opportunity to further this goal. The consequence of his theory-led attitude to the experiment, coupled with his strong political motivation, was that he over-interpreted the data to favor Einstein's theory over Newton's when in fact the data supported no such strong construction. Specifically it is alleged that a sort of data fudging took place when Eddington decided to reject the plates taken by the one instrument (the Greenwich Observatory's Astrographic lens, used at Sobral), whose results tended to support the alternative "Newtonian" prediction of light-bending. Instead the data from the inferior (because of cloud cover) plates taken by Eddington himself at Principe and from the inferior (because of a reduced field of view) 4-inch lens used at Sobral were promoted as confirming the theory. Furthermore Eddington employed a brilliant, as perhaps somewhat misleading, public relations campaign to stampede scientists and the public into accepting his thesis that the somewhat flimsy and suspect data he had obtained amounted to an epochal contribution to science, encompassing a complete overthrow of the Newtonian world system and its replacement by another (Sponsel 2002).

Those who believe that there is no smoke without fire will not be surprised to hear that nearly every factual statement in the preceding narrative, taken in isolation, is basically true, though there are, of course, caveats. However, once the whole story is fully constructed from such documentation as has survived, I believe it is easily seen that the overall picture presented is completely wrong. Specifically there is no direct link, nor does it seem that one can draw a link, between Eddington's self-admitted predisposition to believe the theory and the story of how the critical data came to be selected as it was.

Let us begin with a few points to restore some balance to the picture of Eddington as a master manipulator. Keep in mind that pacifists and draft-resisters like Eddington were a tiny and frequently despised minority during the First World War. Many went to jail, as Eddington was apparently prepared to do himself (Chandrasekhar 1976, p. 250-251). Within the scientific community in allied countries there were few voices willing to be heard in favor of postwar reconciliation in 1919. At the time that Eddington was on Principe the final ratification of the Versailles treaty by the German parliament and government was accompanied by unrest in the streets of Einstein's Berlin. The bitterness left by the war was such that even Einstein and Eddington, two of the war's more prominent and steadfast opponents, feared to write to each other lest they give offence to a former "enemy," until after initial attempts to bring them together by scientists in neutral Holland. When, as a result of the eclipse expedition, Einstein was almost awarded the Gold Medal of the Royal Astronomical Society for 1919/20???, a heavy backlash amongst members of the Society prevented the award being made at all for that year, in spite of Eddington's best efforts to prevent this.

Not only was Eddington swimming against the tide politically, his view that general relativity was correct placed him in a small minority within the astronomy community (and probably also, but to a less extreme extent, within the physics community). Most astronomers were skeptical of, or frankly hostile to, this theory, in so far as they understood it at all, though Einstein's great reputation and the gradual acceptance which special relativity had achieved demanded that the theory be accorded a very respectful reception. The narrative outlined above demands not only that Eddington achieve a monumental impact on the public and scientific mind, but that he do so from a position of profound weakness. Therefore we should proceed with caution in imagining that this miracle could be accomplished entirely through the oratorical brilliance, public-relations savvy and scientific prestige of one (or two) individuals, no matter how great.

The usual view is that Eddington hardly thought the experiment worth performing (Chandrasekhar 1976, p. 250), as he was so sure of the result. There is, indeed the famous anecdote, told by Eddington himself, which encapsulates his theory-led outlook:

"As the problem then presented itself to us, there were three possibilities. There might be no deflection at all; that is to say, light might not be subject to gravitation. There might be a `half-deflection', signifying that light was subject to gravitation, as Newton had suggested, and obeyed the simple Newtonian law. Or there might be a `full deflection', confirming Einstein's instead of Newton's law. I remember Dyson explaining all of this to my companion Cottingham, who gathered the main idea that the bigger the result, the more exciting it would be. `What will it mean if we get double the deflection?' `Then,' said Dyson, `Eddington will go mad, and you will have to come home alone.'" (Chandrasekhar 1976, p. 250)

One observes, in contradistinction to this, that Eddington was conscious of the possibility of an unexpected result as the date of the eclipse drew near, as witnessed by his closing remarks in an article describing the forthcoming expedition.

"It is superfluous to dwell on the uncertainties which beset eclipse observers; the chance of unfavourable weather is the chief but by no means the only apprehension. Nor can we ignore the possibility that some unknown cause of complication will obscure the plain answer to the question propounded. But, if a plain answer is obtained, it is bound to be of great interest. I have sometimes wondered what must have been the feelings of Prof. Michelson when his wonderfully designed experiment failed to detect the expected signs of our velocity through the aether. It seemed that that elusive quantity was bound to be caught at last; but the result was null. Yet now we can see that a positive result would have been a very tame conclusion; and the negative result has started a new stream of knowledge revolutionizing the fundamental concepts of physics. A null result is not necessarily a failure. The present eclipse expeditions may for the first time demonstrate the weight of light; or they may confirm Einstein's weird theory of non-Euclidean space or they may lead to a result of yet more far-reaching consequences – no deflection." (Eddington 1919, p. 122)

It is still noteworthy how theory-led Eddington's viewpoint is. Besides the null result he admits of only two possibilities, the two theoretical predictions made by Einstein at different stages of the development of his theory. I think Eddington was conscious all along of the limitations of the experiment. He and his colleagues would have been utterly over-ambitious to have made very high claims of precision on the order of a repeatable experiment such as Michelson's. Eddington had already shown himself prepared to cast aside Einstein's theory if another one, which might be more in accord with measurement could be found, as shown by his interest in the unified field theory of the Swiss mathematician Hermann Weyl, at a time when Einstein's theory seemed likely to fail the Solar redshift test. One December 16, 1918, only a few months before the planned expedition, Eddington had written to Weyl:

"One reason for my interest [in your paper] is that it seems to me to reopen the whole question of the displacement of Fraunhofer lines, leaving the theoretical prediction unsettled. (Perhaps you will differ from me as to this). [Charles E.] St. John and [John] Evershed seem to be quite decided that experimental evidence is against the deflection, and this is rather a severe blow to those of us who are attracted by the relativity theory. I venture to think your theory may show a way out of the difficulty – but that is a guess." (ETH-Bibliothek Zurich, Hermann Weyl Nachlass, Hs 91:522)

What follows will largely take the form of a response to one of the contentions in a famous 1980 paper by John Earman and Clark Glymour, one of a landmark trio of papers by those authors on the three classical tests of General Relativity. Earman and Glymour's papers raise a number of thoughtful points about the status of the light-bending test conducted during the 1919 eclipse, most important among them that Eddington's careful framing of the theory test as a showdown between Einstein's theory and an earlier prediction of Einstein's which Eddington labeled the Newtonian one, played a critical and, at one time, little noticed role in the acceptance of the theory. They also observe that it was the phenomenal success of Eddington's campaign to win acceptance for the light-bending falsification of Newton's theory that prompted skeptics of the theory, such as the Solar Astrophysicist Charles St. John, to reverse the negative verdict on the third redshift test of the theory which they had previously pronounced. However nothing in their paper had the impact of their suggestion that Eddington may have been motivated by bias in favor of Einstein in throwing out some of the data on dubious grounds. I suspect that the outline of this view (not the particulars) did not originate with Earman and Glymour but may have been circulating orally, in one form or another, amongst physicists for some time before the publication of their paper in 1980. Certainly the story has taken on a life of its own since their paper appeared. Sadly as time has gone by there has been little attempt to follow up on the fact that Earman and Glymour themselves offered up some evidence against the thesis that Eddington's personal bias played a significant role in the data analysis.

The trajectory of this story is also an interesting example of how a sufficiently compelling narrative, widely recounted, can quickly evolve from the carefully phrased version found in a scholarly article to a bare bones version found in popular discourse, stripped of all the caveats which originally modulated its dramatic elements. Earman and Glymour's article was used heavily in an account of the eclipse expedition in Harry Collins' and Trevor Pinch's *The Golem*, a book has been much more widely read than the original article. Collins and Pinch are still very careful in not relying on the claim that Eddington fudged any data, their argument being essentially that there is no such thing as a definitive experiment which resolves all doubts and proves one theory over another. This treatment in turn seems to have served as the principal source for a more popular book by John Waller (Waller 2002), whose aim is to debunk some of the well worn anecdotes of scientific progress which have been scrutinized closely by modern historians of science.

When one finally gets to the reader reviews of Waller's book posted to amazon.com one sees all of the scholarly analysis pared down to the essentials. There it is stated (quoting from several different reviewers, from a version of the website

http://www.amazon.com/gp/product/customer-reviews/0192805673/sr=8-1/qid=1182979869/ref=cm_cr_dp_all_helpful/104-5654845-3967159?ie=UTF8&n=283155&qid=1182979869&sr=8-1#customerReviews accessed on June 27, 2007) that

"Eddington's observations of the eclipse over West Africa in 1919, which supposedly proved Einstein's theory, were worthless. Horror! … It's painful to discover that the guiding stories of one's lifetime are nonsense, but sometimes it just has to be done. This book has done it to me. John, you've broken my heart, but I've come through it stronger and wiser. Alas, I now see the old wives' tales everywhere, so I spend my days shuddering and shaking my head. As the motto of 'The X-Files' put it: Trust No-One."

"And we have pioneers like Robert Millikan and Arthur Eddington who made data fit a chosen theory, rather than the other way around. Yet, far from belittling such men, this book shows them in a new and more human light that transforms our understanding of scientific discovery."

"Remember learning in school how Eddington proved Einstein's theory of relatively by comparing the position of stars during and after an eclipse? Actually his images were so poor they proved precisely nothing except that Eddington was a dab hand at faking results. The book catalogues a series of famous scientists whose passion and belief in a theory blinded them to contrary evidence. In fascinating detail the book describes the circumstances surrounding the experiments both in the laboratory and in the wider social context. What links these scientists is that, as it turns out, the theories they were expounding happened to be right - just not for the reasons they gave. This compelling book should be compulsory reading for all students of science and is delightful food for thought for anyone interested in science."

Thus two main points have come across in the journey from scholary article, through increasingly popular (though scholarly written) books, to the vox populi of the web. They are, firstly, that Eddington fudged, faked, or fit his own results to the theory he believed to be true and, secondly, that one must never take a story at its own evaluation, no matter how plausible it may seem, (in this case, the story that Eddington proved that General Relativity was true).

While I will argue that the specific issue of the first point is quite wrong, I certainly cannot fault the attached moral. Indeed it is true to say that the claim that Eddington's data was much dodgier than most people thought, itself probably arose as a reaction to the wrong-headed belief that the 1919 expedition had somehow "proved" General Relativity all by itself. While one can confidently argue that the eclipse measurements were not a very stringent test of relativity, this early and very compelling narrative has now spawned, in reaction, a distorted myth of its own according to which the eclipse results were poor grounds on which to overthrow Newton's theory and replace it with Einstein's. I will argue that, on the contrary, the eclipse results gave rather good grounds for believing that Einstein's theory was better than Newton's when dealing with the strong gravitational fields close to a massive body like the Sun.

## Origins of the Expedition

To begin with a brief overview of the eclipse expedtions, Eddington was one of two leaders of the endeavor, the acknowledged senior man being Sir Frank Watson Dyson the Astronomer Royal and, in that capacity, director of Britain's leading observatory at Greenwich. It was Dyson who originally observed that the 1919 expedition would be uniquely suited to make this test, since the Sun would be in the star field of the Hyades, the closest open stellar cluster to the Earth (Dyson 1917). The expeditions were organized by the Joint Permanent Eclipse Committee of the Royal Society and the Royal Astronomical Society, a committee which Dyson chaired. Dyson was, at all times, the principal organizer and director of the two expeditions, each of which was staffed by one of two different observatories. Eddington, as director of the Cambridge Observatory, led the expedition to Principe Island off the west coast of Africa, accompanied by Edwin Turner Cottingham, a Northamptonshire clockmaker who maintained many of the instruments at Cambridge. The other expedition, posted to Sobral, in northern Brazil, was mounted by the Greenwich Observatory and consisted of Andrew Claude de la Cherois Crommelin, an assistant at Greenwich who hailed from what was about to become Northern Ireland, and Charles R. Davidson, an experienced computer at Greenwich. The expeditions were devoted exclusively to the test of the light-bending prediction. Although the instrumentation is central to the fudging thesis, I will not discuss it at great length, instead referring readers to Earman and Glymour's excellent article. It does not bear critically on my argument concerning Eddington's culpability and the question of the *motivation* for the alleged fudging. For the moment it suffices to say that Eddington brought one instrument with an "Astrographic lens," chosen for its wide field of view, and borrowed from the Oxford Observatory, and that the Greenwich team at Sobral had two instruments, one employing their own Astrographic lens, the other a 4-inch lens borrowed from the Royal Irish Academy, which may have been brought as a backup instrument because it, alone of the three instruments, had been used on eclipse expeditions previously. Because it is central to modern accusations of bias, I will focus mainly on the data taken with the Sobral Astrographic lens.

For ease of transportation, the astronomers decided no to bring along the telescopes with their complex mountings within which the lenses were normally mounted for observatory work. Instead, the lenses were placed into new tubes fitted with coelostat mirrors at one end. Rather than moving the whole instrument in order to maintain a fixed image of the stars as the Earth turned, only the coelostat mirror would be moved by clockwork during an exposure. After the expeditions Dyson concluded that the coelostat mirrors had performed poorly and should not be used on future expeditions to test the theory. Previous eclipse expeditions had been interested in studying the Sun (a coelostat is usually employed in solar observations), but the 1919 expedition wished to be able to take images of as many stars as possible in the general vicinity of the Sun. Therefore lenses with a wider field of view than those typically employed on eclipse expeditions were vital to the project. Earman and Glymour observed that the only recently concluded First World War greatly restricted the instruments available. The Greenwich observatory had still not recovered the instruments it had sent to a 1914 eclipse in Russia, which had been left behind as a result of the sudden outbreak of the war.

## The Data Analysis

Eddington was concerned that a result be announced as soon as possible after the eclipse. Since it would take weeks to travel back to England, and since at least some time would have to be devoted to onsite preparations and the taking of check or comparison plates (about which more later) he wished that data reduction would begin in the field (Chandrasekhar 1976, p. 253). He urged that both expeditions should bring along a small micrometer to measure star positions. In the event, the Brazilian expedition does not appear to have made any attempts at data reduction until after its return (the probable reasons for this will become apparent later), despite remaining on site for two months in order to take comparison plates of the eclipse star field in the absence of Sunlight. Eddington, on the other hand, began making measurements at once, in spite of bad weather obscuring most of the stars in the field, and in spite of making a comparatively hasty departure for home. He wrote to his mother on board his return ship, giving a glimpse of his uncertainty about his data.

"We developed the photographs 2 ea. night for six nights after the eclipse, and I spent the whole day measuring. The cloudy weather spoilt my plans and I had to treat the measures in a different way from what I intended; consequently I have not been able to make any preliminary announcement of the result. But the one good plate that I measured gave a result agreeing with Einstein and I think I have got a little confirmation from a second plate." (Eddington to his mother, Sarah Ann Eddington, aboard SS. Zaire, June 21, 1919, Trinity College Cambridge Library, Eddington Papers, A4/9)

Eddington arrived home probably sometime in July and there is every reason to believe that he conducted the rest of the data reduction of his plates himself. On the other hand, with only a handful of stars on a few plates to work with, owing to the cloud over Principe, he must have been anxious to learn what was on the Sobral plates, as he was aware that the other expedition had had better luck with the weather. The Sobral expedition arrived back in England on August 25, 1919 (Crommelin 1919b, p. 281). Reduction of the data on their plates probably began almost immediately. Worksheets documenting the measurements made are preserved in the Royal Greenwich Observatory archives (now housed in the University of Cambridge Library). The first page, dealing with the plates taken by the Astrographic lens, is headed "Total Solar Eclipse – 1919 May 28-29- Sobral – Astro No. 1" and dated September 2, 1919 (Cambridge University Library MS.RGO.8/150). Some of the sheets are initialed CD and HF, for Charles Davidson and Herbert Henry Furner, both longtime computers at Greenwich. The following year, Dyson commented, in a letter to the American Geologist Louis Agricola Bauer (1865-1932) on July 1, 1920 (Cambridge University Library MS.RGO.8/147),

"Dear Prof. Bauer,

Your long list of "errata" rather alarmed me, though I could not believe that any serious error had been made in the reduction of the "Einstein' photographs, as both Davidson and I have dealt with some thousands of astronomical photographs in very similar fashion, in fact <almost> identical fashion except for the inclusion of the term $\alpha$ giving the displacement." [The displacement is that due to gravitational light-bending. Note also that the word almost has been deleted by Dyson in the preceding quote.]

This suggests that Dyson was intimately involved in the reduction process, and presumably took the lead in it. Certainly some of the key pages in it appear to be written in his hand. The data reduction proceeded throughout the month of September, while in Germany Einstein himself waited with bated breath (he wrote to his close friend Paul Ehrenfest in Holland on Sept 12 to check whether the Dutch scientists, having closer contacts with their English colleagues, had received any news [Einstein 2004, Doc. 103, p. 154]).

On September 12 Eddington and Cottingham spoke before the British Association for the Advancement of Science at Bournemouth, discussing only briefly the significance of their endeavors for relativity theory. They focused instead on an interesting sidelight of their observations, the enormous solar prominence shown very clearly on the Principe plates and still today the most recognizable feature of the 1919 eclipse. All Eddington had to say concerning the actual results was described in the report on the conference as follows

"Professor Eddington gave an account of the observations which had been made at Principe during the solar eclipse. The main object in view was to observe the displacement (if any) of stars, the light from which passed through the gravitational field of the sun. To establish the existence of such an effect and the determination of its magnitude gives, as is well known, a crucial test of the theory of gravitation enunciated by Einstein. Professor Eddington explained that the observation had been partially vitiated by the presence of clouds, but the plates already measured indicated the existence of a deflection intermediate between the two theoretically possible values 0.87" and 1.75". He hoped that when the measurements were completes the latter figure would prove to be verified. Incidentally Professor Eddington pointed out that the presence of clouds had resulted in a solar prominence being photographed and its history followed in some detail; some very striking photographs were shown." (Eddington and Cottingham 1920, p. 156)

These photographs showing the prominence most clearly were precisely those, taken through the thicker cloud at the beginning of the eclipse, which were utterly useless for testing Einstein's theory.

Thus Eddington gave nothing away, if indeed he had anything to give away, about the state of the Sobral data reduction, but confined himself to his own plates and cagily committed himself only to a value between the two theoretical predictions which Einstein had made. The first, due to Einstein and dating from 1910, was based merely on the principle of equivalence and the old special relativity theory. Eddington had labeled this the "Newtonian" prediction. The second, dated after Einstein finalized his theory in 1915, derived from the complete theory of general relativity, and differed from the earlier prediction because it took account of the curvature of spacetime in a gravitational field, which is a well known feature of that theory.

The result from Principe, as it was later published, is much closer to the larger relativistically "correct" 1915 value, but in view of the limitations of his data Eddington was understandably unwilling to claim too much in the way of accuracy in advance of the Sobral results. Nevertheless he was confident at this stage that light-bending, of some magnitude, was an established fact. He had managed to weigh light. It only remained to check whether spacetime was curved.

On or before (almost certainly just before) October 3, Eddington received the long awaited news from Dyson concerning the reduction of the final plates from Sobral, those from the 4-inch lens, which he refers to as the Cortie plates. These plates are named after Father Aloysius Laurence Cortie, SJ (1859-1926), a Jesuit astronomer who had originally been intended for the party at Sobral, but had to be replaced by Crommelin when his duties prevented him from traveling. The 4-inch lens was loaned to the expedition through the good offices of Father Cortie. Eddington replied to Dyson

"Dear Dyson,

I was very glad to have your letter & measures. I am glad the Cortie plates gave the full deflection not only because of theory, but because I had been worrying over the Principe plates and could not see any possible way of reconciling them with the half deflection.
I thought perhaps I had been rash in adopting my scale from few measures. I have not completed my definite determination of A (5 different Principe v. 5 different Oxford plates), it is not greatly different from the provisional though it reduces my values of the deflection a little." (Eddington to Dyson, October 3, 1919, Cambridge University Library MS.RGO.8/150)

This suggests that the Sobral Astrographic data was reduced first and that Eddington had been informed that the value derived was rather close to Einstein's prediction of 1910, which Eddington did not expect to be confirmed. This had clearly worried Eddington as he felt his own data tended to support the higher result predicted by general relativity. Indeed he seems to have doubted that his results could be made to agree with the lower figure, despite confessing to some efforts in that direction. Thus he greets the news that the Sobral 4-inch results strongly favor the higher value with relief.

There is much here that can be made to fit the model of Eddington as a partial experimenter, but there is one important detail that throws cold water on the overall data-fudging narrative outlined above. That is that this letter strongly implies that Eddington was not involved in the reduction of the Sobral data. The tenor of this opening paragraph indicates that he is receiving his first news of the 4-inch data reduction by letter, and there is no suggestion that he had any earlier input, or even any prior information beyond the cumulative results from the Astrographic instrument. Indeed since receiving that earlier information he has occupied himself not with any analysis of data from Sobral but with a reanalysis of his own data to see whether it can be made to fit. Coupled with Dyson's comment, quoted above, that he himself was responsible for the Sobral data, along with Davidson, we can conclude that the Sobral data reduction was conducted independently of Eddington. The initials on the data sheets, the fact that the reduction was undoubtedly done at Greenwich and not at Cambridge, where Eddington was, and the fact that the eventual report was written with Dyson solely responsible for discussion of the Sobral data and Eddington for Principe, all reinforce this impression.

It seems likely that Eddington was never present at Greenwich during the Sobral data reduction. In a letter to Dyson on October 21, 1919 (Cambridge University Library MS.RGO.8/150) he refers to having acquired a season railway ticket, rather suggesting he had not been traveling up to see Dyson in the preceding weeks or months. We can be fairly confident that Eddington simply was not privy to the reduction of the Sobral data or to the crucial decision, recorded in the data sheets (in what appears to be Dyson's hand), to reject the Astrographic data and accept the 4-inch data as "the result of the Sobral expedition." (Cambridge University Library MS.RGO.8/150) The eclipse expedition appears to be a reasonably good case of independence being preserved between the two wings of the collaboration.

If Eddington did not make the decision to exclude the Astrographic data, it is clear that it must have been Dyson who did so. It therefore turns out to be largely irrelevant for the central data-fudging claim, what Eddington's motivations were. We must ask instead, what was Dyson's attitude to war, peace and relativity? Can we claim that he was biased, or overly influenced by Eddington, a younger but more brilliant colleague?

To get a flavor of a viewpoint which was much more common than Eddington's, let me quote from a letter written by Rudolf Moritz (1878-1940), a London barrister and astronomy enthusiast who had made a special study of relativity theory, to Phillip Herbert Cowell (1870-1950), a former assistant at Greenwich who had a reputation as a formidable computer. He was well known for using his calculation skills (in collaboration with Crommelin) to predict the date of reappearance of Halley's comet in 1910. Cowell had apparently asked Moritz for an explanation of relativity theory (special and general), and Moritz' reply (dated March 1st, 1918) is preserved in the Greenwich manuscript archive (Cambridge University Library MS.RGO.8/123).

"So much for the first theory of relativity. I can follow it all analytically and physically and I believe it is true. The second theory of Einstein in 1914 is far more speculative and I fear only accord with observations will make me accept it. Besides the analysis is too beastly for words. I can well understand the compatriots of Riemann and Christoffel burning Louvain and sinking the Lusitania."

In other words, the atrocity of inventing the tensor-based mechanisms of differential geometry which underpin general relativity is quite on a par, morally speaking, with the most notorious (to Englishmen) war crimes of World War I.

In contrast with those of Moritz and Eddington, Dyson's views seem to have been moderate, and not unrepresentative, either for an astronomer regarding relativity or for an Englishman regarding reconciliation with Germany (though Moritz' views may have been the most representative of English opinion, at least while the war still progressed). That is to say, I suspect he was skeptical but not intransigent on both counts. We have good evidence for the former, and much more important, issue, concerning his views on relativity. Earman and Glymour have already noted that Dyson was skeptical of the theory, that he "thought it too good to be true." (Earman and Glymour 1980, p. 85) After the announcement of the eclipse results Dyson sent copies of plates from the 4-inch to a number of leading astronomers, several of whom replied with polite words about the clarity of the images, even though most of them were at least privately not well disposed towards general relativity. In replying to one correspondent, Frank Schlesinger of the Yale observatory, Dyson writes on March 18, 1920

"We are planning to send an expedition to Christmas Island in 1922; & I hope it may be possible to send one to the Maldives; & that the Australians may do something. Is it likely that there will be an American Expedition? I hope so, in view of the importance of having the point thoroughly settled. The result was contrary to my expectations, but since we obtained it I have tried to understand the Relativity business, & it is certainly very comprehensive, though elusive and difficult." (Cambridge University Library MS.RGO.8/123)

We have confirmation of this from Eddington, who wrote, on August 18, 1920, to the mathematician Hermann Weyl,

"It was Dyson's enthusiasm that got the eclipse expeditions ready to start in spite of very great difficulties. He was at that time very skeptical about the theory though deeply interested in it; and he realized its very great importance." (ETH-Bibilothek Zurich, Hermann Weyl Nachlass, Hs 91:523)

We do know that Dyson held liberal views on the desirability of postwar reconciliation with Germany. Significantly, his obituary in the Observatory says "after the Great War, when international co-operation in science had lapsed to a considerable extent, Dyson played a prominent part in the reconstitution of international scientific co-operation through the International Research Council (now the International Council of Scientific Unions) and in the formation of the International Astronomical Union … Dyson took an important part in the initial deliberations that resulted in the formation of the Union, which owes much to his wise guidance." (Jones 1939, p. 186) Nevertheless, I doubt that he was a crusader for the cause of pacifism, as Eddington was. It is worth observing that the International Astronomical Union, at its birth, did not permit Germany and its allies to enjoy membership in the new body, as is pointed out by Stanley, who places Eddington's internationalist views in context, as representative of a small minority within British society (Stanley 2003, p. 88). A newspaper clipping preserved in the Greenwich Observatory archives actually discusses a German Zeppelin raid which caused some damage to the Observatory, and denounces an attempt in the German press to justify the targeting of a scientific institution on the grounds that Greenwich, through its meteorological observations, was directly participating in the British war effort.

It seems to me likely that Dyson's views on reconciliation were fairly typical of liberal Englishmen, and very distinct from the radical pacifism of either Eddington or Einstein. It is highly implausible that Dyson would have deliberately fudged data in order to support a theory of which he was skeptical, or to advance an anti-war cause. Nor is it particularly credible that Britain's leading astronomer would have been so eager to please Eddington as to risk his reputation by throwing out data just to please his collaborator. Whatever one thinks of the scientific merit of the work of the 1919 expedition, one must discard the compelling claim that the Greenwich team were motivated by a transparent "political" (in whatever sense of the term) agenda on the part of the experimenters. This conclusion accords well with the obvious fact that the written report. Dyson and Eddington made no attempt to gloss over the discarded data. They gave a full and clear account of it, along with the reasons for throwing it out. Even if one disagrees with their arguments and conclusions, they did not behave as if they had evidence of a conscious fraud to conceal.

Of course in any scientific endeavor there comes the moment at which the scientist becomes convinced of the correct result, after which he or she tends to become a partisan for that result, whatever their original bias. So it is amusing that, in his first draft of the report from Sobral, Dyson endeavored to average the results from the two instruments there, having noticed that their average was very close to the Einstein prediction. In his draft of the paper he states "The mean with these weights is 1."83 and is very close to the value required by Einstein's theory." (Manuscript of report, Cambridge University Library MS.RGO.8/150)

Eddington objected

"I do not like the combination of the astrographic with the other Sobral results – particularly because it makes the mean come so near the truth. I do not think it can be justified; the probable errors of both are I think below 0".1 so they are manifestly discordant. If the results are accepted with the weights assigned, the probable error of the mean (judged from their accordance) is about ± 0".20, which certainly does not seem to do justice to the results obtained. I would like to omit the last 5 lines of p.4. It seems arbitrary to combine a result which definitely disagrees with a result which agrees and so obtain still better agreement." (Eddington to Dyson, October 21, 1919, Cambridge University Library MS.RGO.8/150).

There remain many issues raised by Earman and Glymour and others which I will not address, such as the role which Eddington's personal and scientific views did play, (especially in his own reduction of the Principe data) and the matter of Eddington's and Dyson's presentation of their results to the public. But one may safely conclude, on the basis of the documentary evidence, that there are no grounds to believe that the most critical decision of the eclipse data reduction was taken from a biased standpoint. Whether that decision was right or wrong, and whether it had an unwarranted influence in the reception of one of the leading physical theories of the 20th century, are of course quite separate questions, but I will offer some arguments, in the next section suggesting that the decision in favor of Einstein over Newtonian gravity was justified.

## A Problem of Scale

The method of comparing positions of stars on the plates was substantially similar for both expeditions. Plates taken during the eclipse were clipped together with plates taken in nighttime conditions, so that the star positions were as close as possible to each other. One of the two plates would be a reversed image (taken with the use of a mirror as opposed to a lens) so that the images could be compared face to face in this way. A micrometer was then used to measure the separation between the positions of identical stars on the two plates. In this way it would be known how far star A on one of the eclipse plates was positioned away from the same star A on a comparison plate.

In practice the Greenwich team faced the difficulty that both their eclipse and comparison plates were reversed (by the use of the coelostat mirrors in their instruments), and so could not be compared face to face. They made use of a third plate, which they called a scale plate, specially taken of the same field, but not reversed (i.e. direct), which was placed against each of the eclipse and comparison plates in turn. The Cambridge team's comparison plates were taken using a telescope at Oxford (the astrographic lens used on Principe was loaned to the expedition by the Oxford Observatory), and were thus direct, This permitted them be placed face-to-face with the reversed eclipse plates taken via the coelostat mirror on Principe.

At Greenwich the measurements of the Sobral data were made by Charles Davidson, one of the two Sobral Observers, and Herbert Henry Furner, like Davidson a long time computer at Greenwich. Both had been "established computers" there since the mid-1890s. Furner was first appointed an established computer in 1897, having been a supernumerary computer since 1889, according to his obituary (Melotte 1953). Davidson had been an established computer for about the same length of time (both men are listed as such in the Annual Report of the Astronomer Royal in 1900, as given in Maunder 1900, Chapter 5). The data reduction at Greenwich was done under Dyson's overall direction, by his own testimony (Dyson et al 1920).

The Greenwich team had taken able to take comparison plates of the eclipse star field while still in Brazil. This was possible at Sobral, where the eclipse took place in the morning with the Sun relatively low in the sky. Two months later the Sun had moved far enough away from the Hyades so that the star field had risen almost to the same altitude it had during the eclipse before the Sun itself rose in the morning. This permitted nighttime plates of the field to be taken with the same instrument, at the same location. Although the Greenwich team had originally entertained the possibility of leaving without waiting long enough to take comparison plates in situ, they decided it would be best not to do so after experiencing problems with astigmatism in the coelostat mirror used with their astrographic lens (Dyson et al 1920, p. 298).

Taking comparison plates on Principe (where the eclipse took place in the afternoon) would have required the Cambridge team to remain on the island for half a year, a considerable inconvenience for any European scientist. This was perhaps especially true of Eddington, who told his mother he was anxious to be back in England before the end of the strawberry season, writing "I suppose I shall be back about July 10 & shall look forward to the strawberries, which are better than anything they have in the tropics" (Arthur Stanley Eddington to his mother, 21 June 1919 from S. S. Zaire, preserved in Trinity College, Cambridge Library, Eddington Papers, A4/9). The Sobral data thus more easily facilitated a direct comparison between the stars on one of the eclipse plates, with the stars on a comparison plate, whereas Eddington's task was somewhat complicated by the fact that the images were taken by a different instrument from a completely different place on the Earth's surface. Therefore Eddington also took images of a different starfield at both Oxford and Principe, so he could, by comparison of those plates, make sure there were no systematic differences between the Oxford and Principe eclipse field.

Even so, the task of the Greenwich team did not simply consist in measuring the displacement of the star images between the comparison and eclipse plates and concluding that the resulting raw data was the Einstein displacement. Regardless of the amount of light deflection found, there could be additional differences in star positions between the two plates due to three different kinds of misalignment between them. The first would be whether the centers (more generally the origin) of each plate coincided when clamped together. The second concerned the relative orientation of the two plates, either because of a rotation of the instrument, or simply because of the way the plates were clipped together. Finally there might have been a change of scale on the plates, due to some change in the focus or other property of the instrument between the exposures.

If the star field was photographed at a different altitude in the sky then there would be differences in stellar positions on the plate due to differential refraction in the Earth's atmosphere. Differential refraction refers to the fact that images at a different altitude experience differing amounts of refraction by the Earth's atmosphere, thus causing relative shifts in position (for this reason the Sun's image is never precisely circular except when it is at the zenith). Even if everything else about the two plates was identical, the lapse of time between taking eclipse and comparison plates (two months in the case of the Sobral expedition) meant that the Earth was moving in a different direction, relative to the direction to the star field, thus creating differences in stellar aberration between the two plates. Stellar aberration refers to a shift in the apparent position of stars due to the relative motion of the Earth compared to the line of sight towards the star. These last two kinds of changes could be calculated theoretically. Both teams carried out these calculations during their data analysis, but other changes in scale and orientation between the plates, if they occurred, could not be predicted in advance, but had to be measured.

Forunately changes in the plate position and orientation must, in principle, behave in a way which is characteristically different from the purely radial displacement predicted by Einstein for the light-bending effect. Furthermore they could be minimized by careful handling of the plates when clipping them together. The most important of the three changes which had to be determined in order to convert raw measurements from the plates into light-bending results, was the difference in *scale* between the two plates, because a change in scale between the plates could mimic the actual light-deflection displacement. The scale could, however be distinguished from the light-bending deflection, as pointed out by Eddington himself (Eddington 1919, p. 120), because, if measurements were taken with respect to the position of the Sun at the center of each plate, then the light-bending deflection would be greatest for those stars nearest the Sun, whereas the shift in position due to a scale change would be greatest for those stars farthest away from the origin. Of course distinguishing scale from light-bending deflection in this manner would require measurement of a variety of stars at different distances from the Sun. Having an insufficient numbers of stars to measure would make it impossible to distinguish almost any change in the plates from the light deflection. For instance, in the case of the Principe plates, there were so few stars visible that even the orientation could become difficult to distinguish from the deflection, thus causing some plates to be unusable (Dyson et al 1920, p. 321).

In the jargon of the field, there were six plate constants to be solved for each plate pair (3 kinds of possible misalignment times 2 co-ordinates on each plate). The method used was to set up equations that compared the measured displacements between stars on the eclipse and comparison plates to equations based on the various plate constants and the sought-for displacement. Then overdetermination of the plate constants was employed (there are only half-a-dozen plate constants, plus the light-bending factor itself, and 7 or more stars exposed on nearly all plates at Sobral) to derive values for all of the plate constants, including the scale factor and the light-bending factor. If no other changes of scale existed, calculation of the theoretical differential refraction and aberration could remove two of the plate constants and thus permit a better determination of the remaining constants.

As is well known, there were problems with the images taken with the Astrographic lens, because, as was noted by the observers on site, it had lost focus during the eclipse (possibly as a result of the change in temperature common during total eclipses). In addition Eddington claimed that the cloud at Principe might have been actually beneficial in blocking light from the brighter stars, thus making their images on his plates smaller than those on the Sobral plates (Dyson et al 1920, p. 329). Especially when coupled with smearing or streaking of images, overexposure of a plate may result in star images whose size and shape make measurement of the principal star positions, judged from the center of each star image on those plates, problematic. However, it is not clear that Eddington actually verified whether there really was any overexposure in the Sobral plates.

In the end the Greenwich team decided to measure the star positions on the Astrographic plates only in right ascension, and not in declination (i.e. measuring in celestial longitude, east-west, only and not in celestial latitude, or north-south). The positions of the stars relative to the Sun meant that considerably more of the light-bending effect would be measurable in right ascension than in declination. Given that the data on the Astrographic plates was considered "noisy" it seemed ill advised to try to measure in a coordinate in which the sought-for effect (the "signal") would be smaller compared to the noise. However, as we shall see, this may have resulted in an inaccurate determination of the scale, since so much information about the scale constant was lost by excluding the direction in which the scale would have been more visible, because it would have been larger relative to the light-bending displacement. An inaccurate determination of the scale would obviously result in an inaccurate determination of the Einstein displacement, since the two effects are not very dissimilar.

In the end the result obtained from the Sobral Astrographic plates was discordant with the results obtained from the other two instruments (the 4-inch at Sobral and the Principe Astrographic). It is the particular contention of several modern critics that the decision to discount the results from this instrument must have been largely taken because the result was also discordant with the prediction of General Relativity. I have argued, however, that this seems highly unlikely for the team of astronomers at Greenwich, who were far from being convinced that this theory was correct. The decision to measure only in one coordinate for these plates is clear evidence that the Greenwich team was unhappy with the quality of the data they contained. Indeed problems had been noted at the time of the eclipse itself. This is noted in the report itself (Dyson et al, 1920, p. 309) and also more or less contemporaneously with the first news from the eclipse, in an account given by Dyson to a meeting of the Royal Astronomical Society at its June meeting (Fowler 1919).

The Greenwich team went further however, as we learn from their report. They tried different methods of reducing the data in an effort to better understand it and assess its reliability (for instance a least-squares method for finding the plate orientation, which did not substantially alter the result gained from their preferred method of calculating the plate constants using only the brighter stars on the plates, not surprisingly since it adopted the same average change of scale). The most revealing comment made in the report reads as follows,

"The means of the 16 photographs [taken with the Astrographic lens] treated in this manner [i.e. solving for all plate constants, including scale, and the displacement, from the same data] give
$\alpha + 243\,e = +0^r.0435$ [where $\alpha$ is the light-bending displacement and $e$ is the change of scale between the two plates]
or with the value of the scale $+0^r.082$ from the previous table
$\alpha = +0^r.024 = +0".93$ at the limb.
It may be noticed that the change of scale arising from difference of refraction and aberration is $+0^r.020$. If this value of $e$ be taken instead of $+0^r.082$ we obtain
$\alpha = +0^r.039 = +1".52$ at the sun's limb." (Dyson et al 1920, p. 312)

This value is much closer to that recovered from the other two instruments (deflection of $+1."61$ on Principe, $+1."98$ from the Sobral 4-inch). It certainly suggests the possibility that the reason for the discrepancy is an inaccurate determination (and exaggeration) of the change of scale undergone by the Astrographic instrument. This alternative result appears nowhere else in the report, but the mere fact that it is mentioned suggests that the author (which means Dyson for this section of the report, written in his hand in the manuscript of the report [Cambridge University Library MS.RGO.8/150]) attached some significance to it. Furthermore an earlier comment in the paper may be significant in this context

"These changes [in the focus of the astrographic lens during the eclipse] must be attributed to the effect of the sun's heat on the mirror, but it is difficult to say whether this caused a real change of scale in the resulting photographs or merely blurred the images." (Dyson et al 1920, p. 309)

A straightforward interpretation would be that Dyson suspected that the scale value was not accurately determined from the Astrographic data, and he was therefore justified in ignoring any result derived from that data. But what about the possibility that he suspected the 1".52 deflection at the limb was the true result from that data?

There is one other published mention of this 1".52 deflection. It is not in the report itself, but instead in a single page account of the report as it was given orally to the famous joint meeting of the Royal Society and the Royal Astronomical Society, published by Crommelin in *Nature*. He writes:

"This instrument [the Astrographic] supports the Newtonian shift, the element of which is 0.87" at the limb. There is one mode of treatment by which the result comes out in better accord with those of the other instruments. Making the assumption that the bad focus did not alter the scale, and deducing this [scale], from the July plates, the value of the shift becomes 1.52"." (Crommelin 1919, p. 281)

It is noteworthy that the apparently throwaway character of the remark in the published report (2 sentences out of over 40 pages) is contradicted by Crommelin's decision to devote two sentences to it in the mere page available to him in *Nature*. It is especially interesting that both mentions imply that this method of deriving the deflection is very similar, in important respects, to that employed by Eddington in his reduction of the Principe data, which I will now describe in outline.

According to the published report (Dyson et al, 1920, p. 317) Eddington was solely responsible for the reduction of the Principe data, what there was of it. It will be recalled that his comparison plates were taken in Oxford with a different instrumental setup, making a direct plate-to-plate comparison potentially problematic. Therefore check plates of the starfield around Arcturus were taken with both instruments in both places. Although these were originally intended merely as a safeguard against systematic errors arising out of changes in both instrument and location, in the event his method was to take measurements on the check plates in order to calculate the difference in scale between the two instrumental setups. He could then *assume* that the same change in scale applied to those plates as applied to the eclipse field plates taken in both places. In Eddington's words, from the report

"As events turned out the check plates were important for another purpose, viz., to determine the difference of scale at Oxford and Principe. As shown in the report of the Sobral expedition, it is not necessary to know the scale of the eclipse photographs, since the reductions can be arranged so as to eliminate the unknown scale. If, however, a trustworthy scale is known and used in the reductions, the equations for the deflection have considerable greater weight, and the result depends on the measurement of a larger displacement. On surveying the meagre material which the clouds permitted us to obtain, it was evident that we must adopt the latter course; and accordingly the first step was to obtain from the check plates a determination of the scale of the Principe photographs." (Dyson et al., p. 317)

The material available from Principe was meager indeed. Owing to the cloud, which began to clear just as the eclipse was ending, only 2 plates with 5 stars on each were usable. This was insufficient to allow all 6 plate constants to be determined along with the light-bending displacement, and barely sufficient, even if the data for those 5 stars was perfect (which was far from the case) to calculate 4 plate constants and the displacement given theoretical calculations of the differential refraction and stellar aberration. Thus, in Eddington's case the need for an independent determination of the change in scale was severe. But of course mere necessity does not provide any answer to the principal charge made by Earman and Glymour, that there was no justification for throwing out the results of the Greenwich Astrographic, while keeping the poor quality data obtained on Principe by the Oxford Astrographic.

First, let us quote Eddington's own attempt to justify the inclusion of his data:

"… our result [for the light-bending deflection at the limb of the Sun] may be written 1".61±0".30
It will be seen that the error deduced in this way from the residuals is considerably larger than at first seemed likely from the accordance of the four results. Nevertheless the accuracy seems sufficient to give a fairly trustworthy confirmation of **Einstein's** theory, and to render the half-deflection at least very improbable.

38. It remains to consider the question of systematic error. The results obtained with a similar instrument at Sobral are considered to be largely vitiated by systematic errors. What ground then have we – apart from the agreement with the far superior determination with the 4-inch lens at Sobral – for thinking that the present results are more trustworthy?

At first sight everything is in favour of the Sobral astrographic plates. There are 12 stars shown against 5, and the images though far from perfect are probably superior to the Principe images. The multiplicity of plates is less important, since it is mainly a question of systematic error. Against this must be set the fact that the five stars shown on plates W and X [from Principe] include all the most essential stars; stars 3 and 5 give the extreme range of deflection, and there is no great gain in including extra stars which play a passive part. Further, the gain of nearly two extra magnitudes at Sobral must have meant over-exposure for the brighter stars, which happen to be the really important ones and this would tend to accentuate systematic errors [because it is more difficult to tell where the true center of the star lies when it is over-exposed on the plate and thus has, in effect, become a large blob of emulsion], whilst rendering the defects of the images less easily recognized by the measurer. Perhaps, therefore the cloud was not so unkind to us after all.

Another important difference is made by the use of the extraneous determination of scale for the Principe reductions. Granting its validity, it reduces very considerably both accidental and systematic errors. The weight of the determination from the five stars with known scale is more than 50 percent grater than the weight from the 12 stars with unknown scale. Its effect as regards systematic error may be seen as follow. Knowing the scale, the greatest relative deflection to be measured amounts to 1".2 on **Einstein**'s theory; but if the scale is unknown and must be eliminated, this is reduced to 0".67. As we wish to distinguish between the full deflection and the half-deflection, we must take half these quantities. Evidently with poor images it is much more hopeful to look for a difference of 0".6 than for 0".3. It is, of course, impossible to assign any precise limit to the possible systematic error in interpretation of the images by the measurer; but we feel fairly confident that the former figure is well outside possibility." (Dyson et al. 1920, p. 328-329)

This last paragraph is a close paraphrase of the closing part of Eddington's letter to Dyson on October 3 (the first paragraph of which was quoted above). Evidently Eddington's first, and most critical audience, for his claims that the Principe result should be accepted as valid was his collaborator Dyson.

The key words in this part of the paper are "granting the validity" of the extraneous determination of scale. Should we, in hindsight, grant Eddington this validity? About this Eddington himself said:

"The writer must confess to a change of view with regard to the desirability of using an extraneous determination of scale. In considering the programme it had seemed too risky a proceeding, and it was thought that a self-contained determination would receive more confidence. But this opinion has been modified by the very special circumstances at Principe and it is now difficult to see that any valid objection can be brought against the use of the scale.

The temperature at Principe was remarkably uniform and the extreme range probably did not exceed $4^o$ during our visit – including day and night, warm season and cold season. The temperature ranged generally from $77½^o$ to $79½^o$ in the rainy season, and about $1^o$ colder in the cool gravana. All the check plates and eclipse plates were taken within a degree of the same temperature, and there was, of course, no perceptible fall of temperature preceding totality. To avoid any alteration of scale in the daytime the telescope tube and object-glass were shaded from direct solar radiation by a canvas screen; but even this was scarcely necessary, for the clouds before totality provided a still more efficient screen, and the feeble rays which penetrated could not have done any mischief. A heating of the mirror by the sun's rays could scarcely have produced a true alteration of scale though it might have done harm by altering the definition; the cloud protected us from any trouble of this kind. At the Oxford end of the comparison the scale is evidently the same for both sets of plates, since they were both taken at night and intermingled as regards date.

It thus appears that the check plate is legitimately applicable to the eclipse plates. But the method may not be so satisfactory at future eclipses, since the particular circumstances at Principe are not likely to be reproduced." (Dyson et al. 1920, p. 329-330)

That the Greenwich team also underwent a change of heart in this respect is shown by their preparations for the next eclipse, of 1922. Before the expedition departed, Davidson wrote a paper arguing that the independent determination of scale was not only a superior method to that employed in the reduction of the Sobral data, but would be vital for the 1922 eclipse, which lacked the bright stars close to the Sun that had been a unique feature of the 1919 eclipse. It was, indeed, this happy coincidence which had convinced Dyson that the opportunity of testing Einstein's theory in 1919 could not be passed up (Dyson 1917).

"In the Eclipse of 1919, the field of stars was unusually favourable for a determination of the Einstein gravitational displacement of light passing near the Sun – in fact, there is no other field on the Ecliptic with so many bright stars.

In the Eclipse of 1922, if exposures are given of sufficient length to photograph faint stars near the Sun, there is grave danger of the images being drowned in the Corona. The brighter stars which are sure to be photographed are at such distances that the *differential* Einstein effect will be small, with consequent uncertainty in the result.

If, however, one had an independent determination of the scale of the photograph, then two stars, each situated at 1° distance on opposite sides of the Sun, will show an increase in distance of 0".88, a quantity readily measured on good photographs. A scale may be determined by photographs taken the night before or after on a comparison field, as was done by Prof. Eddington in Principe. To this it may be objected that different conditions hold between the day and night observations." (Davidson 1922, p. 224-225)

Many efforts were made in subsequent eclipses, not always successfully, to get an independent measure of the scale without having to assume that it would be the same during the eclipse as during a later nighttime exposure. Most popular was the taking of check plates actually during the eclipse, as advocated by Davidson in his paper, even though this required changing the pointing of the instrument towards a different star field.

Thus, for most subsequent eclipses, an independent determination of scale was employed, precisely because, as noted Davidson, the smaller the observed effect when deprived of stars close to the Sun, the more danger there was in effectively halving the size of the effect to be sought by calculating the scale change from the same data used for the deflection.

It is important to keep in mind that the Sobral 4-inch plates remain highly unusual, even after many subsequent eclipse expeditions to test Einstein's theory. They are almost unique in being taken with a working instrument, in clear weather, with several bright stars relatively close to the Sun. Even so the experimenters were conscious of the difficulties of dealing with a possibly unknown change of scale. Much space is devoted to a discussion showing that the values adopted for the scale are consistent from one image to another across a plate (Dyson et al. 1920, p. 306-309)

In the case of the Principe plates only the presence of unusually bright stars permits any kind of measurement at all, given the cloudy conditions. The fact that, luckily enough, check plates were available, made it possible for Eddington to derive a result which he, at least, was reasonably happy with. I accept, however, that his admitted biases might have made him especially anxious to extract a result from data that another experimenter would have been tempted to discard altogether.

In the case of the Sobral astrographic plates, the excellent conditions were compromised by the poor performance of the instrument. Although the Sobral plates show more stars than Eddington's at Principe, in practice only the 5 brightest are used for the determination of the deflection (Dyson et al. 1920, p. 310-311). Recall also that these star positions are measured only in one co-ordinate, thus halving the amount of information available to correctly determine the scale (on top of the "halving" of reliability claimed by Eddington as a consequence of not having an independent determination of the scale). In the circumstances something might have been done had check plates been available to permit an independent determination of the scale. But note that Eddington's use of the scale change calculated from his check plates presumes that the scale did not change between the time of the eclipse and the taking of check plates. Eddington claims that the extraordinary evenness of temperature in the unusual equatorial conditions of Principe justifies this. It might also be justifiable at Sobral, even though two months passed between eclipse and the taking of the comparison plates. If one assumes the scale did not change (apart from what is theoretically predictable on the basis of differential refraction and aberration) then one obtains, perhaps significantly, a value for the deflection very similar to what Eddington derived from his data.

Does this mean that the 1919 report should have argued that all three instruments favored Einstein's theory over the "Newtonian" result? Undoubtedly the decision to discount the suspect data altogether was the correct one, but the existence of an alternative result is mentioned in the report suggests that the authors attached some significance to it. It certainly constituted evidence that the data was unreliable. The logic of accepting the "Newtonian" value obtained by the main data analysis demanded that a large part of the deflection observed at Sobral by the astrographic was due to a significant change of scale *within the instrument itself*, presumably due to the change of focus during the eclipse. Accepting the lower deflection result had as a corollary that the instrument had preformed unexpectedly and perversely, which would tend to suggest that data taken with it may not have been terribly reliable.

I suspect that the Greenwich team actually came to believe that the 1".52 value was more likely the correct one, on the grounds that, in hindsight, they considered Eddington's data reduction method superior. But in the absence of check plates there was no way to prove that the instrument had experienced no significant change of scale, so they decided to publicly discount the astrographic data. Psychologically, I claim, it must have been significant to them that there was reason to believe that this instrument also might have yielded a result which was inconsistent with the "Newtonian" value, had matters been handled differently.

## The 1979 Re-analysis

Prolonged examination of the 1919 report by Dyson, Eddington and Davidson, the fruits of which are presented in the previous section, led me to consider the possibility that the Sobral astrographic data might actually have supported the full Einstein deflection after all. Although not myself an astronomer I am married to one, and my wife, Julia Kennefick, pointed out that the original plates might be still preserved at the Royal Greenwich Observatory (RGO). Could more be done, she observed, with modern plate-measuring machines and astrometric data-reduction software, than had been possible in the days of merely human computers like Davidson and Furner?

In June, 2003 I made a trip to Cambridge, primarily for the purpose of looking over Eddington's papers at Trinity College and the RGO's manuscript archive, now housed at the Cambridge University Library, but with the idea of also making enquiries about the original plates. To my surprise I learned from Adam Perkins, the curator of the RGO archives at the Cambridge UL that a modern reanalysis of the data had already been done, in1978, to commemorate the Einstein centenary of 1979. He gave me the name of the man who had set the project in motion, Andrew Murray, who had been in charge of the RGO's astrometry team at that time. It turned out that Murray's name was not on the published paper, but happily he had occasion to write a letter to *The Observatory* on the topic 10 years later which I found (*Murray and Wayman 1989*), and this led me to the original paper (*Harvey 1979*). Thanks to Perkins, and to Donald Lynden-Bell, former director of the Cambridge Institute of Astronomy (successor to the Cambridge Observatory), I eventually got in touch with Murray himself, who very kindly corresponded with me in some detail about the process of the 1978 data-reanalysis. It seems clear that it is to Murray himself that we owe thanks for the inspired idea to subject the plates to a modern astrometric analysis.

In 1978 most of the Sobral plates still survived intact (It appears, by that way, that the Principe plates have not survived). One eclipse plate and one comparison plate taken with the 4-inch lens were missing, and one of its eclipse plates was broken. A few of the Astrographic plates were discolored. In general nothing stood in the way of the project. The plate measurements were made by E. D. Clements (known as "Clem") on the Zeiss Ascorecard instrument at the RGO (by then relocated to Herstmonceux Castle in Sussex). The data reduction software was written by Murray and the process of reduction was carried out by Geoffrey M. Harvey, author of the published paper. Both Clements and Harvey were members of Murray's staff at the RGO. The first item of note is that the modern methods had no particular difficulty with the astrographic data, although the images were sufficiently non-circular, on many of the plates from both instruments, to prevent the use of a completely automated plate-measuring machine, such as the RGO's GALAXY. The fact that the Astrographic plates were measured in both declination and right-ascension during the re-analysis does not even call for particular comment in Harvey's paper. Murray comments, in an e-mail to me, written November 22, 2003

"In 1978 the plates were re-measured individually on the Ascorecard machine at Herstmonceux; each image was centered in a square graticule and the (X,Y) co-ordinates were recorded by means of moiré fringe gratings. The reasonable results obtained, particularly for the "inferior" Sobral astrographic images, would seem to indicate that the problem with the 1919 measurements was not so much in the quality of the images, but rather in the reduction method, which relied very heavily on the experimental determination of the scale constant e. The Herstmonceux plate reductions of course included both co-ordinates on each plate, giving a much better separation of the plate scale from the deflection on the eclipse plates."

In his 1979 paper Harvey compares the 1919 results with those he recovered using modern techniques

Gravitational displacement at the Sun's limb in seconds of arc

| Determination | Displacement |
|---|---|
| Predicted from Einstein's Theory | 1.75 |
| 4-inch plates reduced by Dyson *et al.* | 1.98 ± 0.18 |
| 4-inch plates measured on the Zeiss | 1.90 ± 0.11 |
| Astrographic plates reduced by Dyson *et al.* | 0.93 |
| Astrographic plates measured on the Zeiss | 1.55 ± 0.34 |

Harvey comments (p. 198)

"For the 4-inch plates there is no great difference between the value obtained by Dyson et al. and that from the new measurements, but the error has been considerably reduced. For the Astrographic plates, however, a significant improvement has been achieved by the new measurements. Where the previous reduction yielded a value of 0".93 with an unspecified, large error, the new determination is 1".55 ± 0".34. This is still a weak result, but does provide support for that from the 4-inch plates. Combining the two fresh determinations, weighted according to their standard errors, gives 1".87 ± 0".13, a result which is just within one standard error of the predicted value."

Note that the revised 4-inch result places the Einstein prediction even further outside the limits of the error bars for this instrument. In general eclipse expeditions often recovered a value in excess of the GR prediction and were only sporadically successful in wrestling Einstein's prediction within their error-bars. This situation had barely improved even by the time of the last such expeditions in the mid-1970s. It is only with radio telescopes measuring quasars being occulted by the relatively radio-quiet Sun, thus with no need for an eclipse, that the Einstein value has been precisely confirmed (Will 1993). Of course the so-called Newtonian result fares even worse compared to the high value obtained.

It is remarkable that that the alternative value, using the constant scale assumption, for the Sobral astrographic from 1919 happens to have an almost identical value (1."52 compared to 1."55) to that obtained by the modern re-analysis. Is this mere coincidence, or are there grounds for believing that it is more than that? Certainly the modern value casts grave doubt on the 0.93" value obtained by the original team. Murray, in the same e-mail to me, comments

"We have to remember than in those days, the labour of computation was a problem, so short-cuts had to be taken. In particular, the Greenwich astrographic plates were only measured in one co-ordinate (declination). The general philosophy, both at Greenwich and Cambridge, seems to have been to determine the relative plate scale and individual orientations by some means or other, and then apply these to the measured displacements of individual stars to derive the deflection obtained from each of them.

There can always be a problem with trying to determine independently too many plate constants simultaneously in one solution, because of correlations between them due to the actual geometrical distribution of the stars in the field. Presumably there must be some such effect which has affected the scale derived in Table IX [i.e. the derived scale which apparently gives rise to the erroneous result from the Astrographic plates in Dyson et al.]. We should note that, although the deflection is greater in declination, there is a lot of information on the plates scale in the right ascension direction which has been completely ignored."

He adds in a later letter (Nov 27, 2003)

"I can only infer that the inclusion [by Harvey in 1978] of the right ascension measures on the astrographic plates (which were ignored by Dyson et al.) has considerably improved the solution for the deflection, in spite of its smaller effect."

There is a wealth of extra information in the modern analysis. Rather than being obliged to simply compare pairs of images plate by plate, the astrometric software permitted the construction of a database of positions from the comparison plates, against which the positions of the stars on the eclipse plates could be compared. Thus, the displacement of each star could be compared to the position of every other star on every comparison plate, not just to its own position on one or two comparison plates. In addition the plate-measuring machine was able to provide reliable measurements of position in both co-ordinates, for the astrographic plates. This wealth of extra data meant that there was no difficulty in calculating the scale change.

Since it now seems likely that the low "Newtonian" value was due to errors in reducing the data, it seems plausible that the problem lay in an incorrect value for the scale, given that an incorrect value of other plate constants, such as the orientation, would not have affected the final value so much. Therefore it seems likely that the assumption that the scale did not change much between eclipse and comparison plates was the correct one, though obviously it had to remain only an assumption in the context of 1919.

In hindsight, there seem to be excellent grounds for believing that the Greenwich team did make a justifiable assessment of their data in concluding that it definitely supported the prediction of General Relativity over the "Newtonian" prediction. Nonetheless, most astronomers of the day, including Dyson, wished to see the observation repeated at the eclipse of 1922, before treating the case as closed.

It seem, as one might expect, that the teams who took and handled the data knew best after all. But it is hard to stop a good story once it gets going. Because it is fair to say that the eclipse results over the years were never in perfect accord with general relativity's prediction (even given that they definitely were sufficient to falsify the "Newtonian" value), people had begun to question the stock account of the experimentum crucis that vindicated Einstein. Some were not very impressed with the theoretician's swagger with which Eddington and Einstein had greeted the verdict of the observations (I suspect both men were playing up to this image of the cocky and self-assured theorist). Gradually a notion seems to have arisen that there was something highly suspect about the 1919 results, and the new mood is seen in two papers published in 1980, by the philosophers Earman and Glymour, and by the physicist C.W. Francis Everitt. Everitt, like Earman and Glymour, and apparently independently of them, concluded that the astrographic data was excluded largely on grounds of failure to agree with the theoretically predicted value of GR.

"Others again from an astrographic camera at Sobral gave a very-reliable looking measurement of 0.93±0.05 arc-sec – the scaling coefficient must have been wrong, so they were thrown out though the evidence for them was much better than that for the 16.1 ± 0.30 arc-sec measurement at Principe. It is impossible to avoid the impression – indeed Eddington virtually says so – that the experimenters approached their work with a determination to prove Einstein right. Only Eddington's disarming way of spinning a yarn could convinced anyone that here was a good check of General Relativity."

Nevertheless, as I have argued above, I believe that close inspection of the totality of information now available to us since 1979, suggests that the 1919 experimenters probably were justified in concluding that they had at least falsified the lower Newtonian prediction.

So successful was the new counter-myth of bias on the part of Eddington and colleagues that in 1988 Stephen Hawking included the following passage in his famous book A Brief History of Time.

"It is normally very difficult to see this [light-bending] effect, because the light from the sun makes it impossible to observe stars that appear near to the sun in the sky. However, it is possibly to do so during an eclipse of the sun, when the sun's light is blocked out by the moon. Einstein's prodiction of light deflection could not be tested immediately in 1915, because the First World War was in progress, and it was not until 1919 that a British expedition, observing an eclipse from West Africa, showed that light was indeed deflected by the sun, just as predicted by the theory. This proof of a German theory by British scientists was hailed as a great act of reconciliation between the two countries after the war. It is ironic, therefore, that later examination of the photographs taken on that expedition showed the errors were as great as the effect they were trying to measure. Their measurement had been sheer luck, or a case of knowing the result they wanted to get, not an uncommon occurrence in science. The light deflection has, however, been accurately confirmed by a number of later observations. (p. 32)"

In appears that Hawking was aware that widespread suspicions concerning the original data reduction existed. He may have read Earman and Glymour's paper, or Everitt's, or encountered similar stories which I suspect had been circulating orally amongst physicists for some time. I myself have heard physicists claim that the size of the error in the early eclipse expeditions was the same size as the effect itself, which is not a claim made by Earman and Glymour, and not one which, as far as I can tell, is supported by the evidence. Hawking remembered that a reanalysis had been done, which in itself makes him nearly unique, since I have not found even one paper that cites Harvey's publication (the Science Citation Index lists only the Murray and Wayman letter discussed below). It seems the GRO reanalysis team's results were reported at a meeting of the Royal Society in 1979, which might be where Hawking became aware of their efforts.

Knowing that some re-analysis had been undertaken, and recalling the many stories of a dubious data analysis by the original team, he may naturally have jumped to the conclusion, when writing his book nearly a decade later, that the reanalysis must have given birth to the story he remembered. Nothing, of course, could be further from the truth, as far as the reanalysis goes, and it was for this reason that Murray and Patrick A. Wayman of Dunsink Observatory in Dublin (possessors of what remains of the 4-inch instrumentation used at Sobral) wrote to the Observatory in 1989 to object.

"The result from the 4-inch plates were thus confirmed, with a smaller standard error, and even the very low-weight 13-inch plates give a significant result.

The last attempt to measure the deflection by optical observations [as opposed to observations in the radio band] at an eclipse was in 1972, by a team from the University of Texas. The results from that expedition was: Deflection (arcseconds) = $1.66 \pm 0.19$

Conditions were very far from ideal on that occasion, but by comparison, the results from the 1919 eclipse were very respectable. It is completely unjustifiable to dismiss them as Hawking has done. A recent assessment by Will (Was Einstein Right?, p. 78) that '… these expeditions were triumphs for observational astronomy and produced a victory for general relativity …' would seem to be much fairer."

I suspect there are three myths which arose at different epochs concerning this famous experiment. Initially Eddington and Dyson experienced extraordinary success in winning public acceptance for their thesis that the experiment should be read as an experimentum crucis falsifying Newton's theory and vindicating Einstein. This gave birth to a widespread belief that the eclipse observations had proved general relativity correct. In fact eclipse observations never imposed particularly stringent tests on general relativity, and even the fairly wide error bars that existed were not sufficient to always bring GR's prediction within the range of the measured deflection. Once more accurate tests of relativity became possible from the mid-1950s on a backlash began to take hold, at least amongst physicists. It was realized that one ought to be able to do far better and gradually the notion that eclipses were not much of a test of relativity led to a version of the story, heard by myself in the 1980s, and apparently also by Hawking, that in fact the error bars in such tests were as wide as the effect to be measured. This in turn helped give birth to the third and most recent myth, that the results reported were an outright fraud.

I should emphasize at this point that I do not regard the opinion of the physicists as irrelevant because it was orally transmitted. Obviously oral transmission of knowledge plays a critical role in physics and other sciences. Nevertheless, just as we have seen how texts lose nuance as we move from those written by the core group of experts (recall Earman and Glymour (1980)'s paper and Collins and Pinch (1993)'s book), to those written by non-experts (recall comments quoted earlier from amazon.com's website), so the same is true for oral transmission. Orally transmitted knowledge within the core group of experts is perhaps the most nuanced knowledge of all. I would argue that even texts written by the members of the core group themselves cannot completely capture all of the nuance of their expert opinion, if only because the reader may lack the expertise to correctly assess it. But as oral knowledge moves outside the core group, it too loses nuance and accuracy. In this instance I do not feel that the opinion that "the errors were as great as the effect they [the eclipse observers] were trying to measure" can be regarded as having a factual foundation, though it is based on the very reasonable contention that most eclipse teams tended to underestimate their reported errors, probably a reflection of the difficulty of excluding systematic errors when all measurements must be taken during one brief period of a few minutes duration.

Dennis Sciama's book on *The Physical Foundations of General Relativity* (Sciama 1969) is a good example of a text which had, within the physics community, the same kind of influence which Earman and Glymour's papers had on the science studies community (that is to say, the historians, philosophers and sociologists of science). It is a considered opinion of the true place of the 1919 eclipse expeditions in this history of relativity which may be sometimes taken as a warrant for completely dismissing them from that history. Sciama says

"Eddington himself later referred to it as "the most exciting event I recall in my own connection with astronomy." Ironically enough, we shall see that Einstein's prediction has not been verified as decisively as was once believed. Between 1919 and 1966 there have been fewer than thirty [total solar] eclipses, giving altogether a total observing time of not more than about two hours. (The longest possible duration of a total eclipse is about 7 ½ minutes, and such an occasion occurs very seldom). … In fact results have been published for only six eclipses." (Sciama 1969, p. 69-70)

Looking over the results from all these eclipses (including that of 1919), Sciama concludes

"One might suspect that if the observers did not know what value they were `supposed' to obtain, their published results might vary over a greater range than they actually do; there are several cases in astronomy where knowing the `right' answer has led to observed results later shown to be beyond the power of the apparatus to detect." (Sciama 1969, p. 70)

To be sure, Sciama comes quite close here to charging that one might achieve any result from an eclipse measurement of light bending. Indeed it is interesting how the viewpoints of the physicists and the sociologists, such as Collins and Pinch (1993), dovetail here. Admittedly Sciama regards the eclipse measurements as pathological because of their interpretative flexibility, whereas the sociologists see this as typical of much of science. Nevertheless they both have a point. There may very well be a sense in which the position of general relativity was much stronger, in the social sphere, than one would expect for a new theory, and that the position of Newtonian theory was far weaker than it had been for two centuries, and far weaker than many non-experts appreciated. Despite appearances, the old theory of gravity may have been sociologically primed to fall at this moment. I suspect this is true myself. Nevertheless, we must be careful to put this social nexus, which permitted the overthrow of Newtonianism at this historical juncture, in context. In this instance I believe that context is the context of reception, more so than the context of discovery. I doubt that the most critical aspects of the eclipse data reduction were taken in an atmosphere of pro-relativity bias, and I believe that the best evidence suggests that the decisions which were made were done so judiciously, for scientifically defensible reasons. Dyson and his colleagues may have been readier to find in favor of Einstein than they would have been for many another alternative theory of gravity, but this is far from saying that they were looking for excuses to skew their data against Newton. I think they strongly believed that their measurements could not be reconciled with the "Newtonian" prediction. The existence of a viable alternative theory undoubtedly played a significant role in their willingness to unambiguously falsify Newton, but this is a far cry from the accusations of bias which have become fashionable in recent years.

In any case Sciama's real point is that the eclipse measurements, all of them, not just the 1919 ones, do not particularly vindicate general relativity. He doesn't claim that the data does not support Eddington and Dyson's main contention, which is to say that the results falsify the "Newtonian" result. I contend that much of the change in narrative concerning the eclipse is due to this change in focus, from a contemporary one which was excited by a test, the first ever to do so, which falsified Newton while leaving a rival theory standing, to our modern one which sees Einstein's theory as the undefeated theory fending off all rivals. With the benefit of hindsight, looking back from a time when Einstein's theory occupies the position once held by Newton's, we are struck by how threadbare seem the emperor's old suit of clothes. Fortunately for relativity theory it has much less revealing clothes to wear nowadays.

Earman and Glymour's 1980 paper, the first proper historical study of the expedition, raised several interesting points. They observed that there were questions that could be raised about the long forgotten third instrument which had actually agreed with the so-called Newtonian result. This led to a new version of the story, in which Eddington began to emerge almost as a villain, and this version has clearly gained some traction and become quite widespread. There are elements of truth to all these stories, but in my opinion, the closest we can come to the truth on this matter is to say the following. The 1919 eclipse expeditions established clearly that light-bending in a gravitational field was a real effect. This was, after all, their principle goal. As Eddington put it, their mission was to "weigh light." Secondly they showed that of the two predictions made by Einstein, the evidence strongly supported the higher general relativistic value and appeared to falsify the lower. This conclusion was supported by subsequent eclipse expeditions, by radio observation of occulting quasars from the 1970s on, and by the data re-analysis of the Sobral plates in 1978. Nevertheless, to paraphrase Eddington, myths have weight. They are not easily suppressed or replaced, and mere written accounts cannot necessarily hope to halt the spread of a good story. Excessive nuance is generally alien to myth. For instance it seems that "fudging" and "faking" are the popular words which spring to mind when the more nuanced sociological term "interpretive flexibility" is invoked. Should my own account have the extraordinary good fortune to spawn a myth of its own, who knows what will be made out of it?

## Acknowledgements

I would like to thank Diana Buchwald for suggesting that I look into this subject, and for providing every kind of support throughout the research, logistical, financial and, most importantly, moral. The first two kinds of support were provided in her capacity as director of the Einstein Papers Project, who also employed me, in one capacity or another, throughout the research. I also thank her gratefully for carefully reading and critiquing the final manuscript. My colleague Harry Collins also read and commented perceptively upon the manuscript. As stated in the paper I received help from a number of others during my work, especially Adam Perkins of the Cambridge University Library, and Andrew Murray, formerly of the Royal Greenwich Observatory. My research was enormously facilitated by the very generous help of Charlie Johnston of Flint, Michigan, who was able to direct me to the various archives with material relevant to the expedition, based on his own interest on the eclipse, and showed me copies of items he had already located. Jean Eisenstaedt of the Observatoire de Paris kindly showed me copies of the minutes of the Joint Permanent Eclipse Committee, the originals of which are in the Royal Society Archives in London. I would particularly like to thank Adam Perkins of the Cambridge University Library, for granting permission to quote from the manuscripts in the Royal Greenwich Observatory archive, which are state papers of the United Kingdom. I am also grateful to the Master and Fellows of Trinity College, Cambridge for permission to quote from letters of Arthur Stanley Eddington in their possession. The Library of the Eidgenoessische Technische Hochschule in Zurich, in particular Ms. Yvonne Voegeli, very kindly helped me with correspondence of Eddington and Hermann Weyl in their collection.